\documentclass[sigconf]{acmart}
\usepackage{subfigure}
\usepackage{amsmath,bm}

\def\BibTeX{{\rm B\kern-.05em{\sc i\kern-.025em b}\kern-.08emT\kern-.1667em\lower.7ex\hbox{E}\kern-.125emX}}
    
\copyrightyear{2019}
\acmYear{2019}
\setcopyright{acmlicensed}
\acmConference[xxx]{xxx}{xxx, 2019}{xxx}
\acmBooktitle{xxx}
\acmPrice{xxx}
\acmDOI{xx.xxxx/xxx.xxxx}
\acmISBN{xxxx}

\begin{document}

%
\title{MMED: A Multi-domain and Multi-modality Event Dataset}

%
\author{Zhenguo Yang}
\authornote{Corresponding authors}
\email{zhengyang5-c@my.cityu.edu.hk}

\affiliation{%
  \institution{College of Computer Science and Technology, Guangdong University of Technology}
  \city{Guangzhou}
  \state{China}
}
\additionalaffiliation{%
  \institution{Department of Computer Science, City University of Hong Kong}
  \city{Hong Kong}
  \state{China}
}

\author{Zehang Lin}
\email{gdutlin@outlook.com}
\affiliation{%
  \institution{College of Computer Science and Technology, Guangdong University of Technology}
  \city{Guangzhou}
  \state{China}
}

\author{Min Cheng}
\email{min.cheng@huawei.com}
\affiliation{%
  \institution{Huawei Noah's Ark Lab}
  \city{Hong Kong}
  \state{China}
}

\author{Qing Li}
\email{csqli@comp.polyu.edu.hk}
\affiliation{%
  \institution{Department of Computing, The Hong Kong Polytechnic University}
  \city{Hong Kong}
  \state{China}
}

\author{Wenyin Liu}
\authornotemark[1]
\email{liuwy@gdut.edu.cn}
\affiliation{%
  \institution{College of Computer Science and Technology, Guangdong University of Technology}
  \city{Guangzhou}
  \state{China}
}

%
\begin{abstract}
In this work, we construct and release a multi-domain and multi-modality event dataset (MMED), containing 25,165 textual news articles collected from hundreds of news media sites (e.g., Yahoo News, Google News, CNN News.) and 76,516 image posts shared on Flickr social media, which are annotated according to 412 real-world events. The dataset is collected to explore the problem of organizing heterogeneous data contributed by professionals and amateurs in different data domains, and the problem of transferring event knowledge obtained from one data domain to heterogeneous data domain, thus summarizing the data with different contributors. We hope that the release of the MMED dataset can stimulate innovate research on related challenging problems, such as event discovery, cross-modal (event) retrieval, and visual question answering, etc. 
\end{abstract}

%


%
\maketitle

\section{Introduction}
Real-world events can be defined as something happened or happening, which attract people's attention and impact our lives. Instances of events can include festivals, politics, natural disasters, etc. In reality, whenever an event happens, it can be witnessed by different groups of people (e.g., professionals and amateurs) and get attention of the public. Therefore, tremendous data related to the same events may be distributed in different data domains (e.g., online news domain and social media domain), due to the convenience of sharing data in the era of Web 2.0. The data distributed on the Internet platforms are rich in data modalities, such as text, image, and video, etc. Multimodal data in multiple domains have brought new opportunities for people and researchers to ``sense the world''. 

In terms of multi-modality, one of the most popular application is cross-modal retrieval, which aims to bridge the gap between the data modalities. For instance, an image can be a query to match the corresponding texts in a text database, and vice versa. There are quite a few existing datasets, e.g., Wikipedia dataset \cite{Rasi2010}, PASCAL sentences \cite{Rash2010}, NUS-WIDE \cite{Chua2009}, Flickr-30K \cite{Young2014}, MS COCO dataset \cite{Lin2014}, etc. However, the existing datasets are strongly-aligned paired data \cite{Yang2017a, Yang2017b, Yang2019b}, i.e., the textual contents are the exact descriptions of their corresponding images. In reality, the multimodal data may be associated with each other by sharing the same labels, yet they are not trying to describe each other exactly, e.g., news articles on a news media like BBC News reporting a real-world event and images shared by social media users on Flickr that are about the same event. In contrast, we denote such cases as weakly-aligned unpaired data \cite{Yang2019b}. The datasets used for cross-modal retrieval are collected from one data domain usually, which do not take into account the cross-domain relationships. In addition, the datasets usually have a fix number of object categories, which is not as complex as real-world events.

\begin{figure*}
  \centering
  \includegraphics[width=480pt]{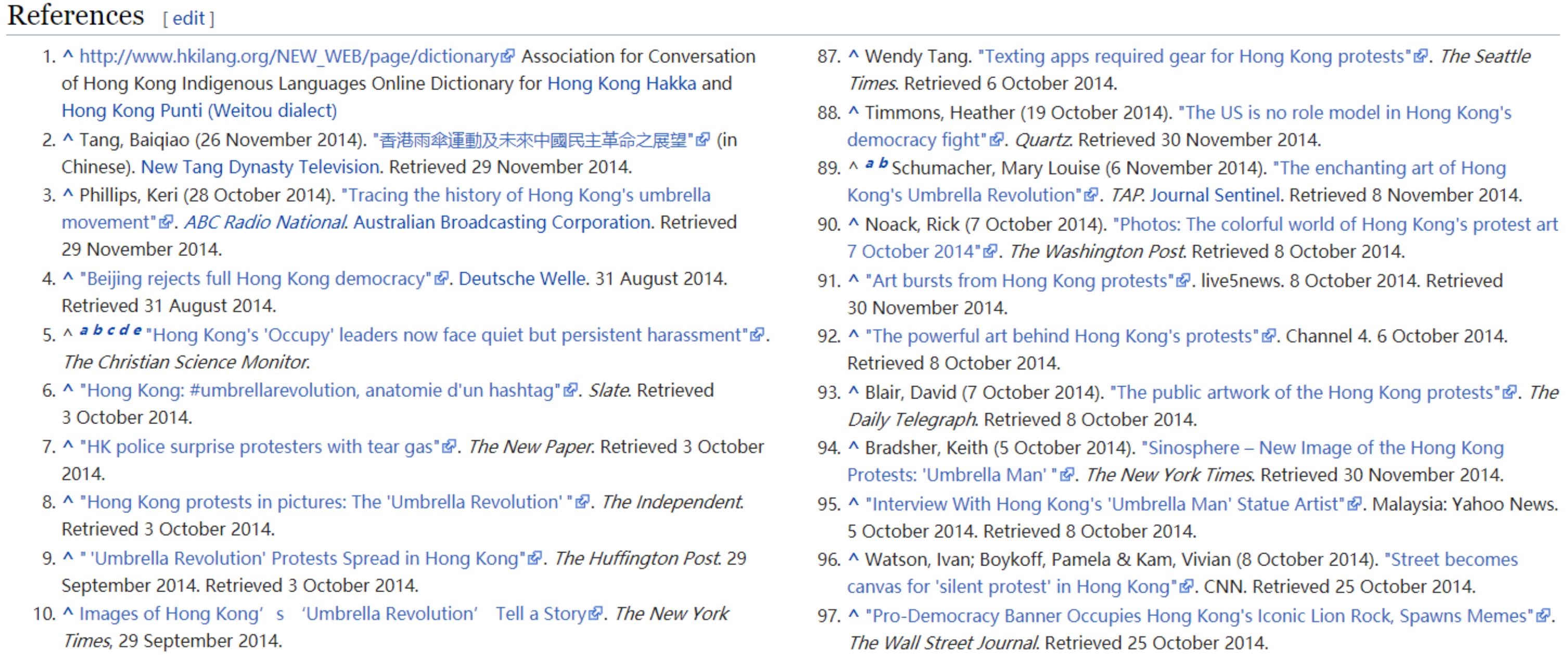}
  \caption{ Snapshot of References of ``Umbrella Movement'' in the Wikipedia Entry.}
\end{figure*}

In terms of events, there exist quite a few works and some datasets for real-world event detection, e.g., event detection from traditional news media \cite{Allan98}, Twitter-like social media \cite{Sakaki2010}, and Flickr-like photo-sharing social media \cite{Papado2011, Yang2017a, Yang2017b}, etc. However, these datasets about real-world events involve one data domain merely. In reality, an influential event happens, the related data may be distributed on different Internet platforms by both amateur users and professional journalist. The rich data provide multiple viewpoints about the real-world event, benefiting to the comprehensive understanding of the events. The distributed data can be complementary or even sometimes contradictory. The single domain data losses the possibility of exploring real-world event related issues crossing data domains.

In this paper, we present a large-scale event dataset collected from multiple data domains in multimodal forms, denoted as MMED, including 25,165 textual news articles from hundreds of news media sites, and 76,516 images from Flickr-like photo-sharing social media. The two-domain data samples are related to 412 real-world events, covering various event categories, such as public security (e.g., shooting, attack, killing, conflict, explosion, epidemic, crash, etc.), natural disaster (e.g. earthquake, flood, fire, etc.), protest, sport, election, festival, etc. Compared to the existing datasets, MMED has several characteristics. Firstly, it is a weakly-aligned unpaired dataset, where the data samples are not trying to describe with each other, which is unlike the strongly-aligned datasets being used for cross-modal retrieval as mentioned previously \cite{Yang2019b}. Secondly, the multimodal data describing real-world event concepts crossing different data domains, and contributed by both amateur users and professional journalists, covering both folk and official viewpoints about events. The dataset has been released on GitHub (https://github.com/zhengyang5/ACM-MMSys19-MMED400). Hopefully, it can promote the research on related applications.

The rest of the paper is organized as follows. Section 2 reviews the principle of data collecting and data annotations. Section 3 introduces the statistics of the dataset. Section 4 sheds light on the possible applications based on the dataset. Section 5 shows the concluding remarks.

\section{Collection and Annotation}
The event labels in MMED cover a wide range of event categories (or event types) like emergency, natural disaster, sport, ceremony, election, protest, military intervention, economic crisis, etc. We collect the dataset considering the aspects of high relevance in supporting the application needs, wide range of event types, non-ambiguity of the event labels, imbalance of the event clusters, and difficulty of discriminating the event labels, etc. 

\begin{figure*}
  \centering
  \includegraphics[width=430pt]{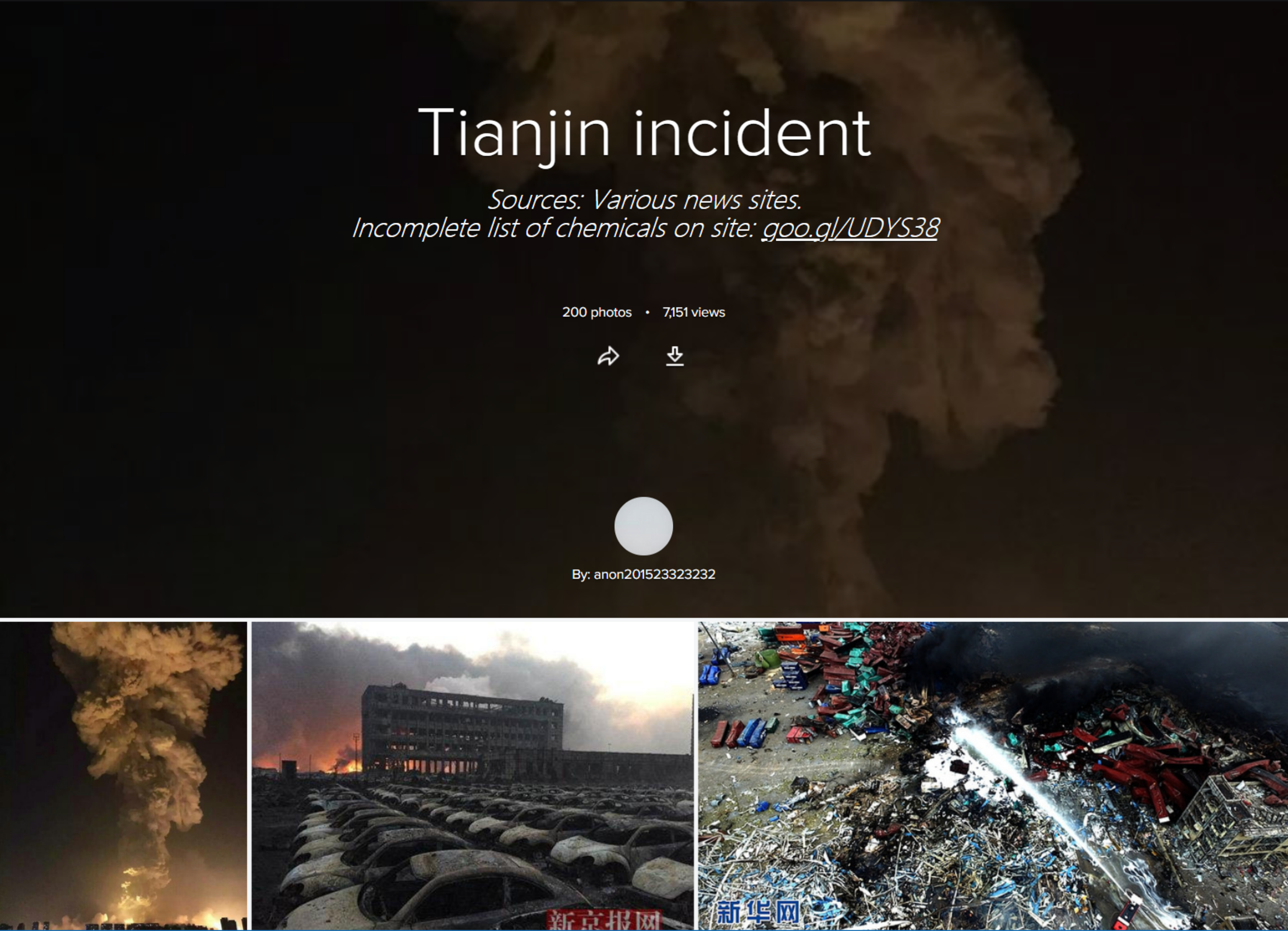}
  \caption{ Snapshot of an Album containing 200 photos about ``Tianjin Explosion'' on Flickr.}
\end{figure*}

To collect textual news articles, there are two challenging problems. The first problem is the dispersibility of the data. For instance, news articles reporting the same events may be reported by any news media sites in the world, thus it is laborious and even impossible to collect the data by accessing each news media sites for annotation. In addition, most online news media sites do not provide effective and convenient interface for retrieval. The second problem is the quality of the data, such as authority, credibility, etc. We have to ensure the news articles are publicly-accepted, and avoid rumors, fake news, or inaccurate news stories as much as possible. Therefore, instead of collecting the data directly from the individual news media sites, we resort to Wikipedia and collect the data in a top-down manner. More specifically, we manually look up the Wikipedia entries about events recently, and check the crowd-sourced articles describing the events that are available online, which have cited quite a few news articles in the references as shown in Figure 1. These articles are contributed by different news media sites, and usually focus on different aspects or hold different views on the events, but are quite high in terms of quality and authority. Furthermore, we crawl these news articles further by accessing the links of the articles in different news media sites. As a result, the name of the Wikipedia entry about the event (i.e., event label) are well-accepted by the public as they are edited in a crowdsourcing manner, and there is no ambiguity in terms of event labels. In addition, Wikipedia provides quite a few portals to access similar events in terms of event types, regions, etc., which helps to collect similar events in certain aspects to ensure the difficulty of discriminating the event labels.

\begin{table*}[h]
  \caption{Statistics of the MMED dataset ($X$: online news media, $Y$: social media).}
  \setlength{\tabcolsep}{2pt}
  \begin{tabular}{|l|l|l|l|l|l|l|l|l|}
\hline
Domain&\#Items&\#Events&Time (\%)&Location (\%)&Title (\%)&Data source/
&Tags (\%)&	Textual content /
\\
 & & & & & &Username (\%)& &Description (\%)\\
\hline
$X$&25,165&412&97.06&n.a.&100&100&n.a.&100\\
\hline
$Y$&76,516&412&100&16.35&100&100	&81.53&100\\
\hline
\end{tabular}
\end{table*}

For the image posts on social media, we take Flickr as an example for collecting data. Given the event names collected from Wikipedia, we retrieve the related data by using different queries (i.e., keywords) and some strategies like filtering by time to obtain a number of returned results and remove the replicated ones. On one hand, we annotate the image posts manually to verify whether the samples are related to the query events, and crawl the responding data samples. On the other hand, we further access the Flickr albums (as shown in Figure 2) of these image posts. Whenever the topics of the albums are in accord with the events, we will craw them further. The event labels of the Flickr image posts are consistent with the ones obtained from Wikipedia to ensure non-ambiguity and difficulty of differentiating.

\section{The MMED Dataset}
\subsection{Statistics of the MMED Dataset}
Specifically, 25,165 textual news articles are collected from hundreds of data sources in the online news media domain, including Yahoo News, Google News, Huffington Post, CNN News, New York Times, NBC News, Fox News, Washington Post, The Guardian, etc. A number of 76,516 Flickr image posts are included, which are shared by 4,473 Flickr social media users. All the data samples are related to 412 real-world events. The number of data samples corresponding to the event labels are summarized in Figure 3.

\begin{figure*}
\centering
\setcounter{subfigure}{0}
\subfigure[Event ID: 1-103]{
\includegraphics[width=480pt]{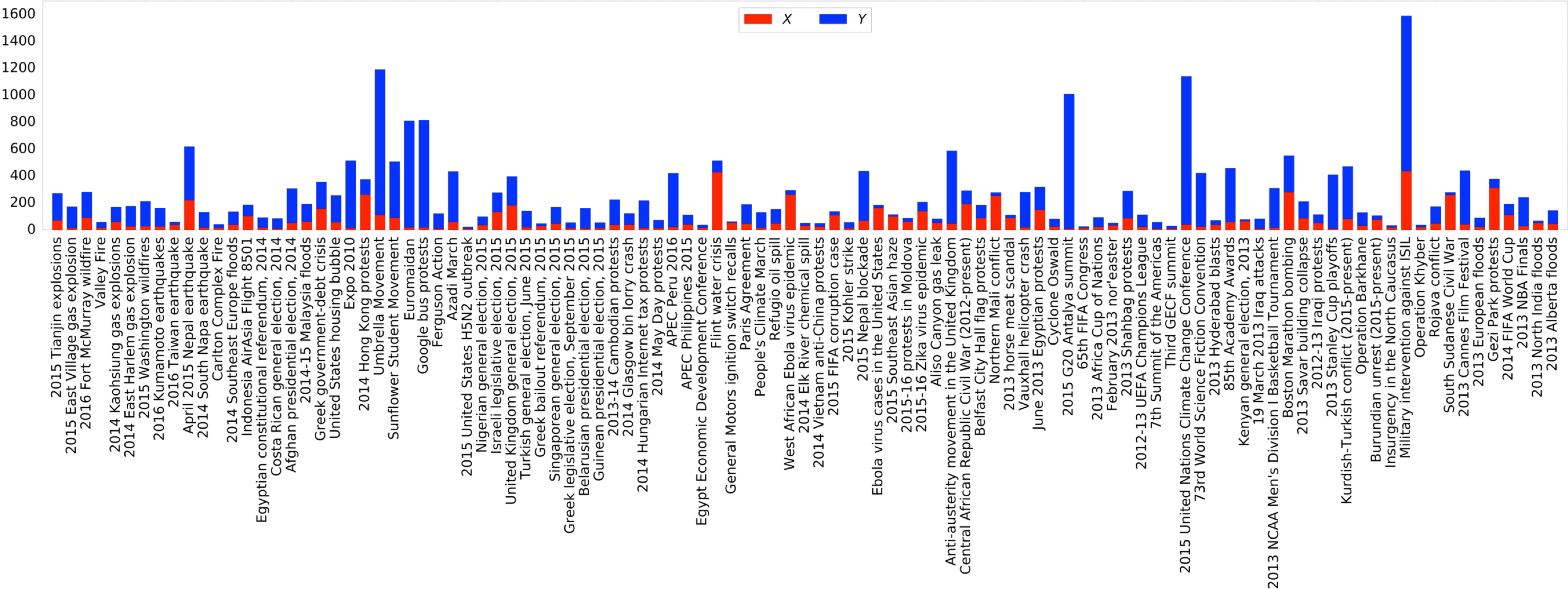}}

\subfigure[Event ID: 104-206]{\includegraphics[width=480pt]{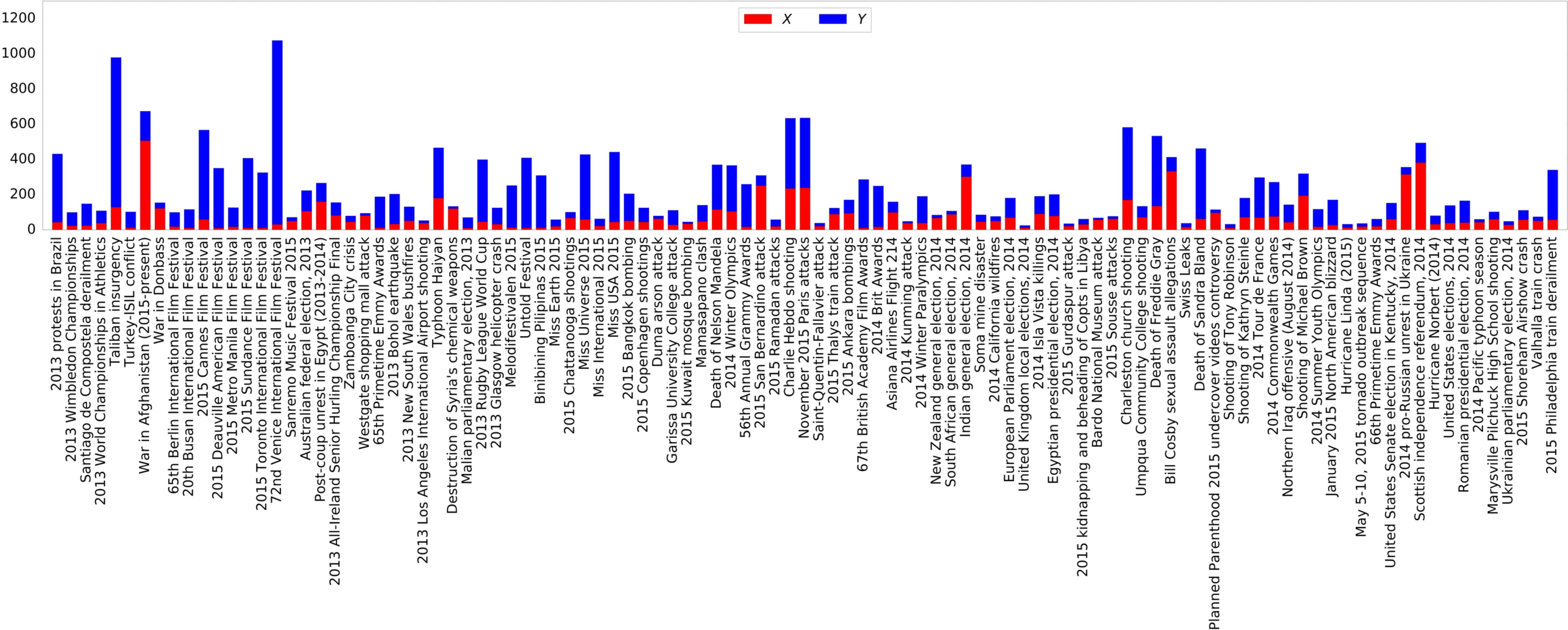}}
\subfigure[Event ID: 207-309]{\includegraphics[width=480pt]{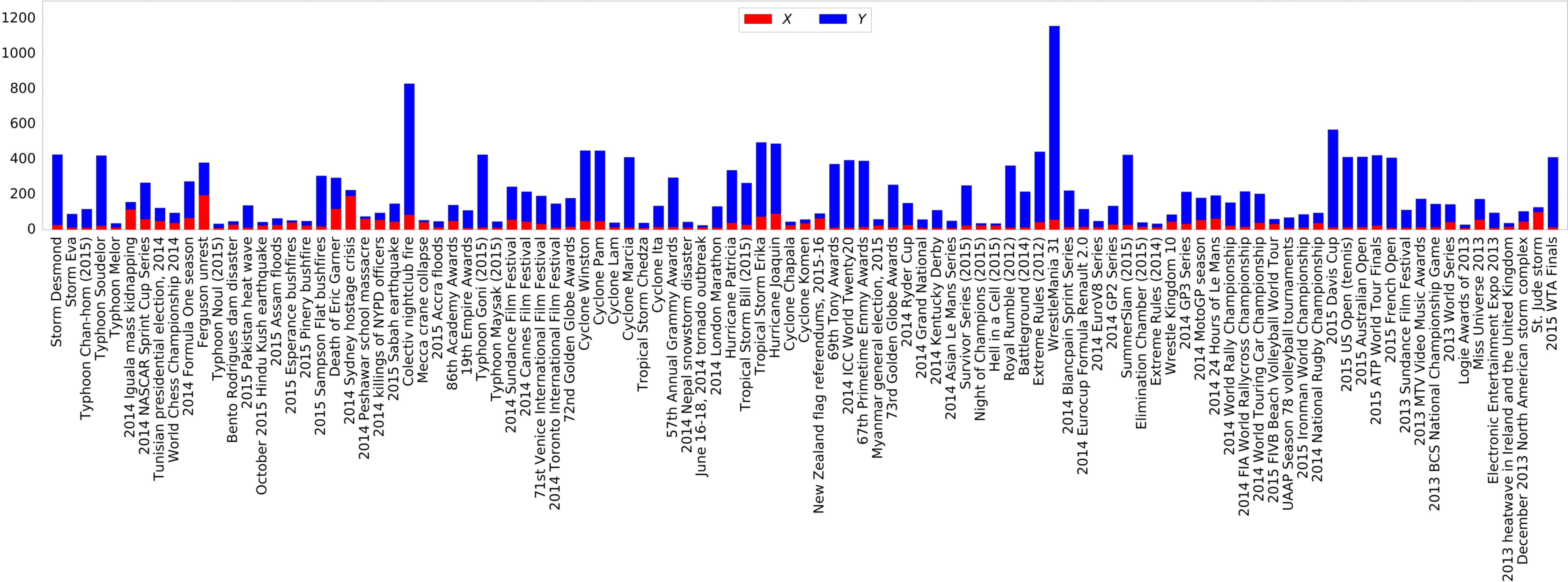}}

\end{figure*}

\begin{figure*}[h]
\centering
\vspace{0pt}

\vspace{0pt}
\subfigure[Event ID: 310-412]{\includegraphics[width=480pt]{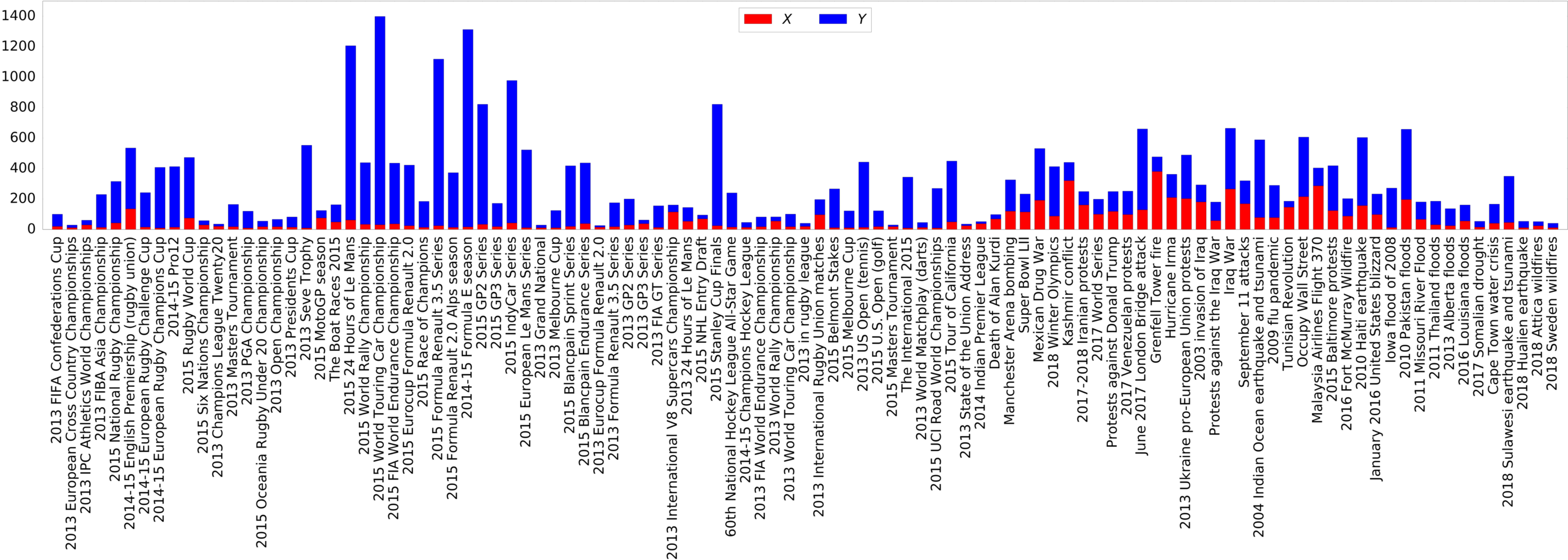}}
\caption{ Distributions of the number of items in each event. (The event names can be referred to the corresponding Wikipedia entries for more information.)}
\end{figure*}

For each textual news article, in addition to its \emph{textual context}, \emph{time reporting}, \emph{title}, \emph{data source} indicating which news media it comes from, are also available. 

For a Flickr image, in addition to the \emph{photo}, \emph{time-taken}, \emph{title}, \emph{GPS-tags}, \emph{tags}, \emph{textual description}, \emph{username}, have been provided. In summary, the statistics of the feature modalities possessed by the data in the dataset are summarized in Table 1.

\subsection{Event Labels of the MMED Dataset}
The real-world events in the dataset can be roughly summarized by several categories, i.e., public security, natural disaster, protest, sport, election, festival, etc. The following Tables summarize the names of event labels and the number of data samples in each domain for the six categories. In particular, there is a Wikipedia entry for the name of each event label, and more details about the events can be obtained by directly accessing these entries in Wikipedia. 

\begin{table}[h]
  \caption{Examples of public security events in the dataset. Note that the numbers indicate the number of data samples in each domain.}
  \setlength{\tabcolsep}{2pt}
  \begin{tabular}{|l|l|l|}
\hline
\bfseries{Event Label}&\bm{$X$}&\bm{$Y$}\\
\hline
2014 Isla Vista killings&81&143\\
\hline
2015 Copenhagen shootings&131&100\\
\hline
2015 San Bernardino attack&269&272\\
\hline
2015 Bangkok bombing&103&68\\
\hline
2015 Tianjin explosions&58&125\\
\hline
...&...&...\\
\hline
\end{tabular}
\end{table}

\begin{table}[h]
  \caption{Examples of natural disaster events in the dataset.}
  \setlength{\tabcolsep}{2pt}
  \begin{tabular}{|l|l|l|}
\hline
\bfseries{Event Label}&\bm{$X$}&\bm{$Y$}\\
\hline
2013 Bohol earthquake&64&189\\
\hline
2015 Nepal blockade&51&58\\
\hline
2015 Sabah earthquake&52&399\\
\hline
April 2015 Nepal earthquake&91&850\\
\hline
Cyclone Oswald&75&254\\
\hline
...&...&...\\
\hline
\end{tabular}
\end{table}

\begin{table}[h]
  \caption{Examples of sport events in the dataset.}
  \setlength{\tabcolsep}{2pt}
  \begin{tabular}{|l|l|l|}
\hline
\bfseries{Event Label}&\bm{$X$}&\bm{$Y$}\\
\hline
2013 Stanley Cup playoffs&62&130\\
\hline
2013 World Series&94&200\\
\hline
2014 24 Hours of Le Mans&164&113\\
\hline
2015 24 Hours of Le Mans&48&107\\
\hline
2015 MotoGP season&58&97\\
\hline
...&...&...\\
\hline
\end{tabular}
\end{table}

\begin{table}[h]
  \caption{Examples of protest events in the dataset.}
  \setlength{\tabcolsep}{2pt}
  \begin{tabular}{|l|l|l|}
\hline
\bfseries{Event Label}&\bm{$X$}&\bm{$Y$}\\
\hline
2013 Shahbag protests&106&83\\
\hline
2014 Hong Kong protests&46&64\\
\hline
Belfast City Hall flag protests&63&73\\
\hline
Boko Haram insurgency&63&80\\
\hline
Euromaidan&105&100\\
\hline
...&...&...\\
\hline
\end{tabular}
\end{table}

\begin{table}[h]
  \caption{Examples of election events in the dataset.}
  \setlength{\tabcolsep}{2pt}
  \begin{tabular}{|l|l|l|}
\hline
\bfseries{Event Label}&\bm{$X$}&\bm{$Y$}\\
\hline
Australian federal election, 2013&69&117\\
\hline
Canadian federal election, 2015&66&369\\
\hline
Egyptian presidential election, 2014&72&112\\
\hline
European Parliament election, 2014&112&122\\
\hline
Indian general election, 2014&190&125\\
\hline
...&...&...\\
\hline
\end{tabular}
\end{table}

\begin{table}[h]
  \caption{Examples of festival events in the dataset.}
  \setlength{\tabcolsep}{2pt}
  \begin{tabular}{|l|l|l|}
\hline
\bfseries{Event Label}&\bm{$X$}&\bm{$Y$}\\
\hline
2014 Sundance Film Festival&75&52\\
\hline
2015 Tour of California&89&399\\
\hline
85th Academy Awards&342&1154\\
\hline
86th Academy Awards&159&67\\
\hline
Miss USA 2015&214&184\\
\hline
...&...&...\\
\hline
\end{tabular}
\end{table}

\section{Application Scenarios and Evaluations}
\begin{figure}[b]
  \centering
  \includegraphics[width=250pt]{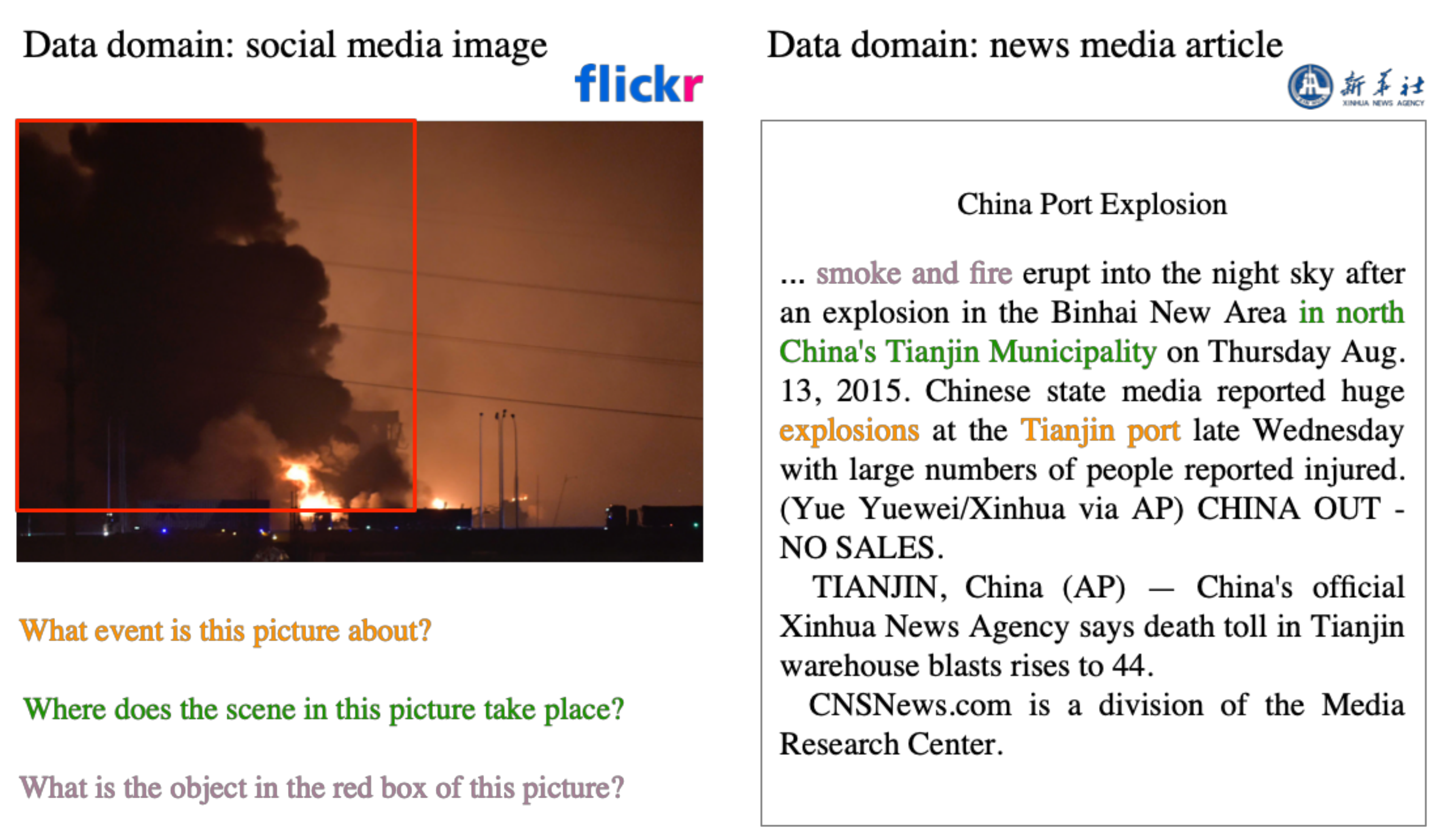}
  \caption{ Example of questions on social media (Flickr) image and answers from news media (Xinhua News) article. Note that the questions in different colors can be answered by the corresponding colored texts from a different domain.}
\end{figure}
\subsection{Event discovery}
Event discovery \cite{Yang2017a, Yang2017b, Yang2019a} aims to detect and organize the data distributed on the Internet platforms based on the real-world events they depict. The collected dataset supports event detection from multiple data domains, such as social media sites and news media sites. The data are in multiple modalities, such as images and texts, and can be contributed by different uses, such as amateur users and professional journalists. In addition, it can be used for event discovery in a transfer learning or zero-shot learning manner, i.e., there are no event labels in the target domain that are available. 

In terms of evaluations, F1-measure NMI can be adopted as the event labels are well-defined.

\subsection{ Cross-modal (event) retrieval}
The scenario of cross-modal retrieval is that using a text query to retrieve the related images. Similarly, we extend the scenario to cross-modal event retrieval \cite{Yang2019b} with two unique characteristics. Firstly, the data are weakly-aligned and unpaired, where the multimodal data are not trying to describe with each other while both are describing high-level semantics like events. In contrast, the existing datasets as mentioned previously are all strongly-aligned and paired data. 

In term of the evaluations, MAP, Precision, Recall and other widely-used metrics in cross-modal retrieval can be adopted similarly.

\subsection{Visual Question Answering}

A VQA system takes as input an image and a free-form, open-ended, natural-language question about the image and produces a natural-language answer as the output \cite{Antoal}. The collected dataset can support the VQA related applications, especially for real-world events. Figure 4 gives an example of questions on a Flickr image, whiles the answers can be found from news media articles. Actually, cross-modal event retrieval tries to answer the first question in the example. 

In terms of evaluations, the performance depends on the distance or similarity between the answers of the methods and the groundtruth, which can be measured by BLEU, ROUGE, METEOR, etc. In addition, user studies on Amazon Mechanical Turk platform can be adopted for evaluations.

\section{Conclusion}
In this paper, we have released an event dataset from social media and news media platforms, denoted as MMED. MMED can be used for three scenarios at least, i.e., event discovery, cross model (event) retrieval, and visual question answering, etc. The real-world dataset is well-labeled, can be used for other scenarios with an open mind.
%
\bibliographystyle{ACM-Reference-Format}
\bibliography{Multimedia}

\end{document}